\documentclass[aps,prc,twocolumn,superscriptaddress,amsmath,amssymb,floatfix]{revtex4-2}

\usepackage{graphicx}
\usepackage{bm}

\newcommand{\Tr}{\mathsf{T}}
\newcommand{\J}{\mathcal{J}}
\newcommand{\Gc}{\mathcal{G}}
\newcommand{\Lc}{\mathbf{H}}

\newcommand{\K}{\mathcal{K}}
\newcommand{\Sc}{\mathcal{S}}
\newcommand{\Id}{\mathbf{I}}
\newcommand{\Vres}{\mathcal{V}_{\mathrm{res}}}
\newcommand{\hbm}{\frac{\hbar^2}{2m}}

\begin{document}

\title{Spurious center-of-mass mode as an $S$-matrix pole within the Jost--random-phase approximation}

\author{Kazuhito Mizuyama}
\email{mizukazu147@gmail.com}
\affiliation{Institute of Research and Development, Duy Tan University, Da Nang 550000, Vietnam}
\affiliation{Faculty of Natural Sciences, Duy Tan University, Da Nang 550000, Vietnam}

\date{\today}

\begin{abstract}
Within the Jost--random-phase approximation (Jost-RPA), the spurious center-of-mass mode is a zero of the Jost
determinant and hence a pole of the $S$ matrix. A zero at zero excitation energy $E$, whether from self-consistency or
from a calibration of the residual interaction, is a pole of exactly second order when it has a single independent
solution, every unoccupied bound orbital lies above the holes, and each hole is paired with every allowed partial wave;
the order follows from the position of the zero, not from self-consistency. The state of this double pole, given by the
Autonne--Takagi factorization of its $E^{-2}$ coefficient, is the zero-energy solution normalized by its inertia. In a
self-consistent model of one energy density of the neutron and proton densities, every physical $S$-matrix element
continued to $E=0$ carries the $E^{-2}$ term, contrary to an earlier conclusion. Subtracting the mode's principal part from the
response changes the strength at real $E\neq0$ only through the smoothing and, in such a model, coincides with the
projection of the ideal spurious state. In uncalibrated Woods--Saxon models of $^{16}$O and $^{208}$Pb the mode is a pair
of poles on the imaginary axis.
\end{abstract}

\maketitle

\emph{Introduction.}\quad In the random-phase approximation (RPA) on a self-consistent mean field, the translation of the nucleus is a solution at
zero excitation energy $E$, the spurious center-of-mass (c.m.) mode~\cite{Thouless1961_RPA}. It is not an intrinsic excitation,
yet it enters the response to $L=1$ external fields. In practice, discrete
RPA and its extensions give a finite real or an imaginary
eigenvalue~\cite{Gambacurta2010_SRPAspur,Papakonstantinou2014_imagSpur,YoshidaNakatsukasa2013_imagSpur,Bianco2014_imagSpur,Knapp2023_spurFinite}.
Continuum calculations place it on the real axis as a bound state of the Green
function~\cite{Agrawal2003_SpuriousMixing}; when it is imaginary, only that fact has been
reported~\cite{Agrawal2003_SpuriousMixing,Khan2002_cQRPA,Papakonstantinou2006_CRPAimagSpur,HamamotoSagawa2002_ISD}. Remedies
rescale the residual interaction to restore zero
energy~\cite{ShlomoBertsch1975,Matsuo2001_cHFBresponse,HaginoSagawa2001_cQRPA,Goriely2004_QRPAalpha,Mizuyama2009_cQRPA},
remove the c.m. coordinate from the external
field~\cite{VanGiaiSagawa1981_NPA,HarakehDieperink1981_ISGDR,HamamotoSagawa2002_ISD,Papakonstantinou2006_CRPAimagSpur},
project the ideal spurious state out of the response~\cite{MizuyamaColo2012_SpuriousSubtraction}, or remove its
coupling from the RPA matrices~\cite{MeyerTerVehn1979}.

Is the mode a pole of the $S$ matrix? Ginocchio, Schucan, and Weidenm\"uller derived the RPA $S$ matrix of
nucleon--nucleus scattering and answered no~\cite{GinocchioSchucanWeidenmueller1970}. Finite-Fermi-system theory treats
c.m. oscillations in an external field as poles of the vertex function~\cite{KhodelSaperstein1982}, and continuum RPA with
a separable force fits the coupling so that a dispersion function vanishes at $E=0$~\cite{Vertse1995_ExactGR}. In discrete
RPA a zero mode with nonzero inertia forms a Jordan pair~\cite{ThoulessValatin1962,Nakada2016_RPAsolutions,Nakada2017_NGJordan},
which gives a factor $E^2$ in the RPA determinant and an $E^{-2}$ term in the
response~\cite{ShimizuDonatiBroglia2000_NGpole,Hinohara2015_NGinertia}, approximately also in a semimicroscopic continuum model~\cite{Gorelik2021_PHDOM}. These concern the response, whereas the 1970 question concerns the physical $S$-matrix elements. As an $S$-matrix
pole the mode has been shown only with the separable-force fit, never as the double pole of a self-consistent model or
as the imaginary pair, whose energy has not been reported.

This Letter answers yes within the Jost-RPA~\cite{Mizuyama2023_JRPA1,Mizuyama2024_Eigenphase,Mizuyama2025_JRPA2}, which
keeps the scattering boundary conditions of every channel and finds the $S$-matrix poles among the zeros of the Jost
determinant: the spurious mode is such a zero. At $E=0$, whether from self-consistency or from a calibration of the
residual interaction, it is, with one state, a pole of exactly second order, and in a self-consistent model every
physical $S$-matrix element continued to $E=0$ has this pole, contrary to Ref.~\cite{GinocchioSchucanWeidenmueller1970}.
In the Woods--Saxon model used here, which approximates a self-consistent theory, the mode is a pair on
the imaginary axis. Subtracting the principal part of the response function at these poles, its Mittag-Leffler terms,
changes the strength at real $E\neq0$ only through the smoothing; in a self-consistent model it equals the projection of
Ref.~\cite{MizuyamaColo2012_SpuriousSubtraction}, and for an imaginary pair it needs no calibration. Pole expansions of
$L=1$ responses and $S$ matrices must contain these poles, whose terms carry $90$--$95\%$ of the energy-weighted sum
rule of the field $r$ here.

\emph{Jost-RPA.}\quad At multipolarity $L$, each of the $M$ particle-hole configurations
$\alpha$~\cite{Mizuyama2023_JRPA1,Mizuyama2024_Eigenphase,Mizuyama2025_JRPA2} pairs a hole $h_\alpha$ of isospin
$q_\alpha$, angular momenta $l_{h_\alpha}j_{h_\alpha}$, and energy $\epsilon_{h_\alpha}<0$ with a particle partial wave
$l_\alpha j_\alpha$ and gives a forward channel $\alpha$ and a backward channel $\alpha+M$, with $l_{\alpha+M}=l_\alpha$
and the wave numbers $k_\alpha(E)$, $k_{\alpha+M}(E)$ of the nucleon energies $\epsilon_{h_\alpha}+E$,
$\epsilon_{h_\alpha}-E$. The $2M\times2M$ Green function, regular at $r=0$ and outgoing at large $r$, obeys
\begin{align}
&\bigl[E\,\tau_z-\Lc\bigr]\Gc(r,r';E)=\Id\,\delta(r-r'),\label{eq:Green}\\
&\Lc=\begin{pmatrix}\mathcal H+\Vres(r)&\Vres(r)\\ \Vres(r)&\mathcal H+\Vres(r)\end{pmatrix},\label{eq:Lc}\\
&\mathcal H_\alpha=-\hbm\Bigl[\frac{d^2}{dr^2}-\frac{l_\alpha(l_\alpha+1)}{r^2}\Bigr]+U^{(q_\alpha)}_{l_\alpha j_\alpha}(r)-\epsilon_{h_\alpha},
\label{eq:Hsp}
\end{align}
with the nucleon mass $m$, the unit matrix $\Id$, the Pauli matrices $\tau_z$, $\tau_x$ on the forward and backward
blocks, $\mathcal H=\mathrm{diag}(\mathcal H_1,\dots,\mathcal H_M)$, and the mean field $U^{(q)}_{lj}(r)$ of isospin $q$ in
the partial wave $lj$; the radial hole functions solve
$\{-(\hbar^2/2m)[d^2/dr^2-l_{h_\alpha}(l_{h_\alpha}+1)/r^2]+U^{(q_\alpha)}_{l_{h_\alpha}j_{h_\alpha}}(r)\}\varphi_{h_\alpha}(r)=\epsilon_{h_\alpha}\varphi_{h_\alpha}(r)$. The residual interaction
and the response function are
\begin{align}
&\Vres(r)=\sum_{qq'}\tilde\varphi_q(r)\,\frac{\kappa_{qq'}(r)}{r^2}\,\tilde\varphi_{q'}^{\Tr}(r),\label{eq:Vres}\\
&R_{qq'}(r,r';E)=\xi_q^{\Tr}(r)\Gc(r,r';E)\xi_{q'}(r'),\label{eq:R}
\end{align}
with the real contact strengths $\kappa_{qq'}(r)=\kappa_{q'q}(r)$, the spherical harmonics $Y_L$, the vector
$\tilde\varphi_q(r)$ whose component $\alpha$ is $\varphi_{h_\alpha}(r)$ times
$\langle l_\alpha j_\alpha\|Y_L\|l_{h_\alpha}j_{h_\alpha}\rangle/\sqrt{2L+1}$ for $q_\alpha=q$ and $0$ otherwise, and $\xi_q^{\Tr}(r)=\bigl(\tilde\varphi_q^{\Tr}(r),\tilde\varphi_q^{\Tr}(r)\bigr)$.

The regular solution $\Phi^{(r)}(r;E)$ and the Jost solutions $\Phi^{(\pm)}(r;E)$ solve $[E\tau_z-\Lc]\Phi(r;E)=0$ with,
for all $2M$ channels $a$, $a'$,
\begin{align}
&\Phi^{(r)}_{aa'}(r;E)\sim\delta_{aa'}\,r\,\mathrm{j}_{l_a}(k_a(E) r)\quad(r\to0),\label{eq:bcr}\\
&\Phi^{(\pm)}_{aa'}(r;E)\sim\delta_{aa'}\,r\,h^{(\pm)}_{l_a}(k_a(E) r)\quad(r\to\infty),\label{eq:bcpm}
\end{align}
where $\mathrm{j}_l$ and $n_l$ are the spherical Bessel and Neumann functions and $h^{(\pm)}_l=\mathrm{j}_l\pm i\,n_l$.
The mean field and the residual interaction vanish beyond a radius, where each column of $\Phi^{(\pm)}(r;E)$ is the Hankel
function of its own channel alone; this fixes the solutions and the Jost matrices $\J^{(\pm)}(E)$ defined by
\begin{align}
\Phi^{(r)\Tr}(r;E)={}&\tfrac12\bigl[\J^{(+)}(E)\Phi^{(-)\Tr}(r;E)\nonumber\\
&+\J^{(-)}(E)\Phi^{(+)\Tr}(r;E)\bigr].
\label{eq:Jost}
\end{align}
With the step function $\theta$, the diagonal matrix $\K(E)$ of the $k_a(E)$, $0\le\arg k_a^{1/2}(E)\le\pi/2$ in
$\K^{1/2}(E)$, and $\Phi^{(\pm)}_F(r;E)=\Phi^{(\pm)}(r;E)\K^{1/2}(E)$, the flux-adjusted $S$ matrix and the Green function
are
\begin{align}
\Sc(E)={}&\K^{\frac12}(E)\bigl[\J^{(+)}(E)\bigr]^{-1}\J^{(-)}(E)\,\K^{-\frac12}(E),\label{eq:Smat}\\
\Gc(r,r';E)={}&-i\frac{m}{\hbar^2}\Phi^{(+)}_F(r;E)\Sc(E)\Phi^{(+)\Tr}_F(r';E)\nonumber\\
&-i\frac{m}{\hbar^2}\bigl[\theta(r-r')\Phi^{(+)}_F(r;E)\Phi^{(-)\Tr}_F(r';E)\nonumber\\
&+\theta(r'-r)\Phi^{(-)}_F(r;E)\Phi^{(+)\Tr}_F(r';E)\bigr].
\label{eq:GS}
\end{align}

Throughout, $E$ lies on the first Riemann sheet, where $\mathrm{Im}\,k_a(E)\ge0$ in every channel, with equality only on
the cuts $|E|\ge\min_\alpha|\epsilon_{h_\alpha}|$ of the real axis; off the cuts, $\Phi^{(+)}(r;E)$ decays and
$\Phi^{(-)}(r;E)$ grows. On the upper side of the cut at $E>0$ the open-channel block of $\Sc(E)$ is the physical $S$
matrix. At a zero $E_g$ of $\det\J^{(+)}(E)$ off the cuts, a common nonzero left null vector of $\J^{(+)}(E_g)$ and
$\J^{(-)}(E_g)$ would, by Eq.~(\ref{eq:Jost}), give a vanishing combination of the columns of $\Phi^{(r)}(r;E_g)$, which are
independent by Eq.~(\ref{eq:bcr}); so the zero is a pole of $\Sc(E)$ and, since $\Phi^{(\pm)}(r;E)$ are analytic there
and only the first term of Eq.~(\ref{eq:GS}) contains $\Sc(E)$, a pole of $\Gc(r,r';E)$, as a function of $r$ and $r'$, of the same order. Conversely, a solution regular at $r=0$ that decays is $\Phi^{(r)}(r;E)v$ with
$\J^{(+)\Tr}(E)v=0$ by Eq.~(\ref{eq:Jost}); a pole has one state if all such solutions at it are multiples of one. The Wronskian
of $\Phi^{(r)}(r;E)$ with itself vanishes, so Eq.~(\ref{eq:Jost}) and the Wronskians of the Jost solutions give
$\Sc^{\Tr}(E)=\Sc(E)$, and at a simple pole $E_g$ with one state the Autonne--Takagi factorization defines the state,
\begin{align}
&\mathrm{Res}\bigl[\Sc,E_g\bigr]=\sigma_g\,\omega_g\,\omega_g^{\Tr},\qquad\sigma_g>0,\quad\omega_g^\dagger\omega_g=1,\label{eq:Takagi}\\
&\mathrm{Res}\bigl[\Gc(r,r'),E_g\bigr]=\psi_g(r)\,\psi_g^{\Tr}(r'),\nonumber\\
&\psi_g(r)=\sqrt{\frac{m\sigma_g}{i\hbar^2}}\,\Phi^{(+)}_F(r;E_g)\,\omega_g;
\label{eq:psig}
\end{align}
at a pole of higher order the coefficient matrix of the most negative power of $E-E_g$ replaces the residue matrix.
Since $k_\alpha(-E)=k_{\alpha+M}(E)$, $\tau_x\Lc\tau_x=\Lc$, and $\tau_x\tau_z\tau_x=-\tau_z$, the matrix
$\tau_x\Phi(r;E)\tau_x$ solves the equation at $-E$ with Eqs.~(\ref{eq:bcr}), (\ref{eq:bcpm}), and Eq.~(\ref{eq:Jost}) gives
\begin{equation}
\J^{(+)}(-E)=\tau_x\,\J^{(+)}(E)\,\tau_x ,
\label{eq:even}
\end{equation}
so $\det\J^{(+)}(E)$ is even in $E$: zeros come in pairs $\pm E_g$, as found for the RPA $S$ matrix in Ref.~\cite{GinocchioSchucanWeidenmueller1970}, and a zero at $E=0$ has even order.

\emph{The spurious mode as a zero of the Jost determinant.}\quad For $L=1$, a zero at $E=0$ with one state is double, even
without self-consistency, under four premises: natural parity, so that $l_\alpha=l_{h_\alpha}\pm1$; every
occupied orbital is a hole; (S1) every unoccupied bound orbital lies strictly above the holes of its isospin; and (S2) the
configurations of each hole contain every allowed $l_\alpha j_\alpha$, occupied ones included. For a channel
function $P(r)$, let $Q_P(r)$ be the solution of $\mathcal HQ_P=P$ that is regular at $r=0$ and decays. The rigid shift of the
occupied orbitals has the channel function
\begin{equation}
P_{T,\alpha}(r)=D_\alpha\bigl[\tilde\varphi_{q_\alpha}(r)\bigr]_\alpha,
\label{eq:PT}
\end{equation}
with $[\ ]_\alpha$ the component of configuration $\alpha$ and $D_\alpha=d/dr+[l_{h_\alpha}(l_{h_\alpha}+1)-l_\alpha(l_\alpha+1)]/(2r)$. By (S2) its transition density is
$\tilde\varphi_q^{\Tr}(r)P_T(r)=\frac{r^2}{2}\partial_r\rho_q(r)$, with $\rho_q(r)$ the ground-state density of isospin $q$.

A pair off $E=0$:\quad Under (S1), (S2) the zeros of $\det\J^{(+)}(E)$ off the cuts lie on the real or the imaginary
axis, and a negative restoring energy of the rigid shift gives an imaginary pair. Let $E_g\neq0$ be a zero
off the cuts and $\psi_g(r)$ any nonzero solution of $[E_g\tau_z-\Lc]\psi_g=0$ that is regular at $r=0$ and decays,
i.e., $\Phi^{(r)}(r;E_g)v$ with $\J^{(+)\Tr}(E_g)v=0$. Let $\psi_{g,X}$, $\psi_{g,Y}$ be its forward and backward parts
and $P=\psi_{g,X}+\psi_{g,Y}$. The difference of the two rows gives $\psi_{g,X}-\psi_{g,Y}=E_gQ_P(r)$, and their sum
then gives
\begin{equation}
\bigl(\mathcal H+2\Vres(r)\bigr)P(r)=E_g^2\,Q_P(r).
\label{eq:EsqP}
\end{equation}
Multiplying by $P^\dagger=(\mathcal HQ_P)^\dagger$ and integrating gives
\begin{equation}
\int_0^\infty\!dr\,P^\dagger\bigl(\mathcal H+2\Vres\bigr)P=E_g^2\int_0^\infty\!dr\,Q_P^\dagger\mathcal HQ_P.
\label{eq:Esq}
\end{equation}
Both integrals are real, as $\mathcal H$ and $\Vres$ are real symmetric and there are no surface terms. The following
pair argument shows that the last integral is positive unless $E_g$ is real, so $E_g^2$ is real. An occupied orbital $h'$ of isospin
$q_\alpha$ and partial wave $l_\alpha j_\alpha$ is an eigenfunction of $\mathcal H_\alpha$ with eigenvalue
$\epsilon_{h'}-\epsilon_{h_\alpha}$. Unless $E_g=\pm(\epsilon_{h'}-\epsilon_{h_\alpha})$, which is real,
Eq.~(\ref{eq:EsqP}) and $\mathcal HQ_P=P$ give the $h'$ component
\begin{equation}
\int_0^\infty\!dr\,\varphi_{h'}[Q_P]_\alpha=\frac{-2\int_0^\infty dr\,\varphi_{h'}[\Vres P]_\alpha}{(\epsilon_{h'}-\epsilon_{h_\alpha})^2-E_g^2}.
\label{eq:pair}
\end{equation} As $\mathcal H_\alpha$ is
symmetric, this component contributes $\epsilon_{h'}-\epsilon_{h_\alpha}$ times its squared modulus. By (S2) the configuration $\alpha'$ of the hole $h'$ and the partial wave $l_{h_\alpha}j_{h_\alpha}$ exists.
The component of $[Q_P]_{\alpha'}$ along $h_\alpha$, with the opposite eigenvalue $\epsilon_{h_\alpha}-\epsilon_{h'}$ of
$\mathcal H_{\alpha'}$, has the same modulus: both contain the same radial integral, and the squared reduced matrix elements of $Y_L$ are
symmetric. These contributions thus cancel in pairs. By (S1) the part orthogonal to the occupied orbitals contributes
positively unless it vanishes. If it vanished, $Q_P$ and $P=\mathcal HQ_P$ would lie along occupied orbitals. The
contributions of $P$ to the transition density $\tilde\varphi_q^{\Tr}P$ would cancel in the same pairs, $\Vres$
would not act, and the projected equations would give $P=Q_P=0$.

On functions whose occupied-orbital components cancel in pairs likewise, only the parts orthogonal to the occupied
orbitals, where $\mathcal H$ is positive by (S1), enter Eq.~(\ref{eq:Esq}). On them Eq.~(\ref{eq:EsqP}) is a Hermitian
problem for $E_g^2$ with the scalar product $\int_0^\infty dr\,P^{\Tr}(r)Q_{P'}(r)$ and a continuum from
$\min_\alpha\epsilon_{h_\alpha}^2>0$. Such a function is $P_T$, as
$\int_0^\infty dr\,\varphi_{h'}D_\alpha\varphi_{h_\alpha}=-\int_0^\infty dr\,\varphi_{h_\alpha}D_{\alpha'}\varphi_{h'}$.
Hence, as in discrete RPA~\cite{Thouless1961_RPA}, the lowest eigenvalue obeys
\begin{equation}
E_g^2\le\frac{\int_0^\infty dr\,P_T^{\Tr}(r)\bigl(\mathcal H+2\Vres(r)\bigr)P_T(r)}{\int_0^\infty dr\,P_T^{\Tr}(r)Q_{P_T}(r)},
\label{eq:bound}
\end{equation}
so a negative restoring energy, the numerator, gives a discrete $E_g^2<0$, an imaginary pair~\cite{Thouless1960_Stability}.

\begin{figure}[tb]
\includegraphics[width=\columnwidth]{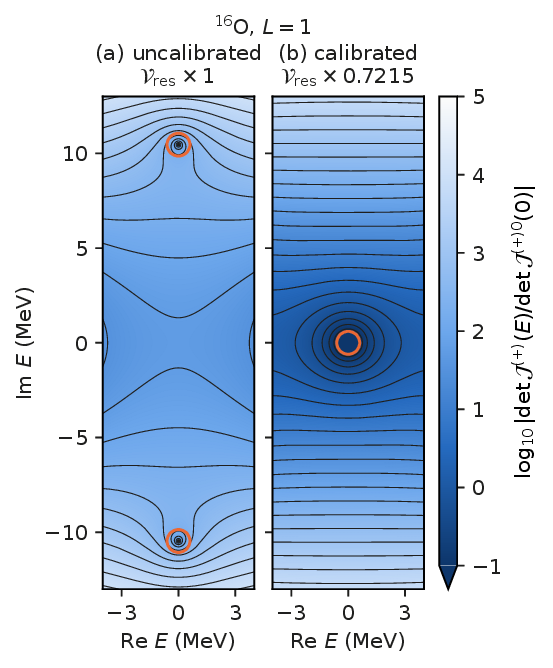}
\caption{$\log_{10}|\det\J^{(+)}(E)/\det\J^{(+)0}(0)|$ for $^{16}$O, $L=1$, on the first Riemann sheet at $|\mathrm{Re}\,E|\le4$~MeV, with contours every $0.25$;
$\J^{(+)0}(E)$ is the Jost matrix of Eq.~(\ref{eq:Jost}) with $\Vres=0$. (a) Uncalibrated Woods--Saxon model: the pair of
zeros at $E=\pm10.46\,i$~MeV (circles). (b) Calibrated model ($\Vres$ multiplied by $0.7215$): the zero of order two at $E=0$ (circle).}
\label{fig:map}
\end{figure}

The Woods--Saxon model used here~\cite{Mizuyama2023_JRPA1} is such a case. Its mean field is
\begin{align}
U^{(q)}_{lj}(r)={}&\Bigl(1\mp0.67\,\frac{N-Z}{A}\Bigr)\biggl\{\frac{V_0}{1+\exp[(r-R_{\rm WS})/a]}\nonumber\\
&+V_{ls}\Bigl[j(j+1)-l(l+1)-\frac34\Bigr]\nonumber\\
&\times\frac1r\frac{d}{dr}\frac{1}{1+\exp[(r-R_{\rm WS})/a]}\biggr\}+\delta_{qp}V_C(r),
\label{eq:WS}
\end{align}
with the upper sign for neutrons and the lower for protons, the neutron and proton numbers $N$ and $Z$, the mass number $A=N+Z$,
$R_{\rm WS}=r_0(A-1)^{1/3}$, and $V_C(r)$ the Coulomb potential of a uniformly charged sphere of $Z$ protons and radius
$R_{\rm WS}$. The depth $V_0$, spin-orbit strength $V_{ls}$, radius parameter $r_0$ and diffuseness $a$ are those of
Ref.~\cite{Mizuyama2023_JRPA1} for $^{16}$O and, for $^{208}$Pb, $V_0=-51$~MeV, $V_{ls}=0.44\,|V_0|\,r_0^2/2=18.10$~MeV\,fm$^2$,
$r_0=1.27$~fm, $a=0.67$~fm~\cite{BohrMottelson1969}. The contact strengths $\kappa_{qq'}(r)$, Eqs.~(59) and (60) of
Ref.~\cite{Mizuyama2023_JRPA1}, are the second derivatives in $\rho_q$, $\rho_{q'}$, at the ground-state densities, of the
$t_0$ term and the $t_3$ term (density exponent 1) of a Skyrme energy density, with $t_0=-1100$~MeV\,fm$^3$,
$t_3=16000$~MeV\,fm$^6$, and the spin-exchange parameters $x_0=0.5$, $x_3=1$; $\hbar^2/2m=20.7213$~MeV\,fm$^2$; the radial step
is $0.2$~fm ($^{16}$O) and $0.25$~fm ($^{208}$Pb). At $20$~fm the Jost solutions take the free Hankel values of
Eq.~(\ref{eq:bcpm}) in every channel, protons included, and are integrated inward with all potentials, Coulomb
included.
For both nuclei the model satisfies (S1) and (S2). The pair lies at $E_g=10.46\,i$~MeV ($^{16}$O) [Fig.~\ref{fig:map}(a)] and
$8.07\,i$~MeV ($^{208}$Pb), with $E_g^2$ below the bounds $-33.3$ and $-35.9$~MeV$^2$ of
Eq.~(\ref{eq:bound}).

A zero at $E=0$:\quad In a self-consistent model the rigid shift gives a decaying zero-energy solution, so
$\det\J^{(+)}(0)=0$. Self-consistency means that the mean field $U^{(q)}_{lj}(r)=U^{(q)}(r)$ and the contact strengths are the first and
second derivatives of one energy density of $\rho_n$, $\rho_p$ and vanish beyond the radius. The hole equation gives
$\mathcal H_\alpha P_{T,\alpha}(r)=-\partial_rU^{(q_\alpha)}(r)[\tilde\varphi_{q_\alpha}(r)]_\alpha$, which
$2[\Vres P_T]_\alpha$ cancels by the transition density of $P_T$ and
$\partial_rU^{(q)}=\sum_{q'}\kappa_{qq'}\partial_r\rho_{q'}$, so $(\mathcal H+2\Vres)P_T=0$. Beyond the radius,
$\varphi_{h_\alpha}(r)$ is a multiple of $r\,h^{(+)}_{l_{h_\alpha}}(k_\alpha(0)r)$, as $k_\alpha(0)=k_{\alpha+M}(0)$
belongs to $\epsilon_{h_\alpha}$. The Hankel-function recurrences give, for $l_\alpha=l_{h_\alpha}\pm1$,
\begin{equation}
D_\alpha\,r\,h^{(+)}_{l_{h_\alpha}}(k_\alpha(0)r)=\mp k_\alpha(0)\,r\,h^{(+)}_{l_\alpha}(k_\alpha(0)r),
\label{eq:DH}
\end{equation}
so $\frac12\bigl(P_T^{\Tr}(r),P_T^{\Tr}(r)\bigr)^{\Tr}$, regular at $r=0$, decays. By Eq.~(\ref{eq:Jost}) this solution is
$\Phi^{(r)}(r;0)v$ with $v\neq0$ and
\begin{equation}
\J^{(+)\Tr}(0)\,v=0,\qquad \det\J^{(+)}(0)=0 .
\label{eq:zeromode}
\end{equation}

In a model that is not self-consistent, a calibration multiplies the residual interaction by a positive constant chosen
so that $\det\J^{(+)}(0)=0$; for the Woods--Saxon model it is $0.7215$ ($^{16}$O) and $0.6116$ ($^{208}$Pb)
[Fig.~\ref{fig:map}(b)].
As $y$ then doubles from $0.05$ and $0.25$~MeV, $\det\J^{(+)}(iy)/\det\J^{(+)0}(iy)$ grows by $4.000$ ($^{16}$O) and
$3.971$ ($^{208}$Pb); with the nonzero $E^{-2}$ term of $R_{qq'}$ below, this shows one state. In the self-consistent
and the calibrated models the zero is, like every zero off the cuts, a pole of $\Sc(E)$. Let it have one state,
with a real solution $\psi_0(r)$, as $\Lc$ is real. Then no $\mathcal H_\alpha$ has a zero-energy bound
solution. By (S1) such a solution would be an occupied orbital $h'$ of the partial wave $l_\alpha j_\alpha$ with
$\epsilon_{h'}=\epsilon_{h_\alpha}$. With $\alpha'$ the configuration of $h'$ and $l_{h_\alpha}j_{h_\alpha}$ in the pair
argument, $\alpha'\neq\alpha$ as $l_{h'}=l_\alpha\neq l_{h_\alpha}$. The vector whose only nonzero components are
$[\tilde\varphi_{q_\alpha}(r)]_{\alpha'}$ in channel $\alpha$ and $-[\tilde\varphi_{q_\alpha}(r)]_\alpha$ in channel
$\alpha'+M$ has zero transition density and, by Eq.~(\ref{eq:Vres}), solves the equation at $E=0$. It and its image
under $\tau_x$ are two independent solutions, contrary to one state. Hence the forward and backward parts of
$\psi_0(r)$, whose difference $\mathcal H$ maps to zero at $E=0$, are equal:
$\psi_0(r)=\frac12\bigl(P_0^{\Tr}(r),P_0^{\Tr}(r)\bigr)^{\Tr}$ with $(\mathcal H+2\Vres)P_0=0$; $P_0=P_T$ in the
self-consistent model.

Exactly second order:\quad With one state, the pole at $E=0$ and the zero of $\det\J^{(+)}(E)$ are of order two in both
the self-consistent and the calibrated models. Since $\det\J^{(+)}(E)$ is not identically zero, the pole of $\Gc(r,r';E)$ at $E=0$ has a finite order $\nu$.
Then $\lim_{E\to0}E^\nu\Gc(r,r';E)=C\,\psi_0(r)\psi_0^{\Tr}(r')$ with $C\neq0$, since its columns are zero-energy
solutions regular at $r=0$ that decay and $\Gc^{\Tr}(r,r';E)=\Gc(r',r;E)$ by Eq.~(\ref{eq:GS}) and $\Sc^{\Tr}=\Sc$. As $\Lc$ maps
$\frac12\bigl(Q_{P_0}^{\Tr}(r),-Q_{P_0}^{\Tr}(r)\bigr)^{\Tr}$ to $\tau_z\psi_0(r)$, multiplying Eq.~(\ref{eq:Green})
from the left by $\psi_0^{\Tr}(r)$ and by the transpose of that vector, and integrating over $r$, gives
\begin{align}
\frac12\int_0^\infty\!dr\,\bigl(Q_{P_0}^{\Tr},Q_{P_0}^{\Tr}\bigr)\Gc(r,r';E)={}&\frac{\bigl(Q_{P_0}^{\Tr}(r'),-Q_{P_0}^{\Tr}(r')\bigr)}{2E}\nonumber\\
&+\frac{\psi_0^{\Tr}(r')}{E^2}.
\label{eq:id1}
\end{align}
The inertia $\int_0^\infty dr\,P_0^{\Tr}Q_{P_0}=\int_0^\infty dr\,Q_{P_0}^{\Tr}\mathcal HQ_{P_0}$ is positive by the pair
argument at $E_g=0$, as no $\mathcal H_\alpha$ has a zero-energy bound solution. Times $E^\nu$, the left side tends
to $\frac C2\int_0^\infty dr\,P_0^{\Tr}Q_{P_0}\,\psi_0^{\Tr}(r')\neq0$, while the right side diverges for $\nu=1$ and
vanishes for $\nu>2$. Hence $\nu=2$ and
\begin{equation}
\lim_{E\to0}E^2\,\Gc(r,r';E)=\frac{2\,\psi_0(r)\,\psi_0^{\Tr}(r')}{\int_0^\infty dr\,P_0^{\Tr}(r)Q_{P_0}(r)}.
\label{eq:G2c}
\end{equation}
In the self-consistent model the hole equation gives $Q_{P_T}$ the components $-(m/\hbar^2)\,r\,[\tilde\varphi_{q_\alpha}(r)]_\alpha$, so $\int_0^\infty dr\,P_T^{\Tr}Q_{P_T}=3mA/(8\pi\hbar^2)$, the c.m.\ inertia. With one state the null space of $\J^{(+)\Tr}(0)$ is
one-dimensional, so some $(2M-1)\times(2M-1)$ minor of $\J^{(+)}(0)$ is nonzero. With no common left null vector of
$\J^{(\pm)}(0)$, the zero of $\det\J^{(+)}(E)$, an Evans-type determinant~\cite{Gesztesy2007_EvansJost}, has the pole's order, two, as in Ref.~\cite{PegoWarchall2002}.

The state of the double pole:\quad The state is $\psi_0(r)$, in the self-consistent model the translation solution,
normalized by the inertia. The coefficient matrix of
$E^{-2}$ replaces the residue matrix in Eqs.~(\ref{eq:Takagi}), (\ref{eq:psig}). It has rank
one by Eqs.~(\ref{eq:GS}), (\ref{eq:G2c}), and its Autonne--Takagi factorization gives
\begin{align}
&\lim_{E\to0}E^2\,\Sc(E)=\sigma_g\,\omega_g\,\omega_g^{\Tr},\nonumber\\
&\psi_g(r)=\pm\Bigl[\frac{2}{\int_0^\infty dr\,P_0^{\Tr}Q_{P_0}}\Bigr]^{1/2}\psi_0(r).
\label{eq:Takagi0}
\end{align}

Physical elements:\quad In the self-consistent model every physical $S$-matrix element, continued to $E=0$, carries this
$E^{-2}$ term. The element $\Sc^{\rm ph}_{\alpha\beta}(E)$ between forward channels $\alpha$, $\beta$ is
$\Sc_{\alpha\beta}(E)$ at real $E$ above both thresholds. Under (S1), (S2) no zero of $\det\J^{(+)}(E)$ lies in the
quadrant $\mathrm{Re}\,E>0$, $\mathrm{Im}\,E>0$, so its continuation through that quadrant to $E$ near $0$ is
$\Sc_{\alpha\beta}(E)$:
\begin{equation}
\lim_{E\to0}E^2\,\Sc^{\rm ph}_{\alpha\beta}(E)=\sigma_g\,\omega_{g,\alpha}\,\omega_{g,\beta}.
\label{eq:Slim}
\end{equation}
Column $\alpha$ of $\Phi^{(+)}_F(r;0)$ is $k_\alpha^{1/2}(0)\,r\,h^{(+)}_{l_\alpha}(k_\alpha(0)r)$ beyond the radius,
so, with $P_0=P_T$, $\omega_{g,\alpha}$ is fixed by the tail of $P_{T,\alpha}(r)$ and is nonzero in every forward channel by
Eq.~(\ref{eq:DH}) and (S2).

\begin{figure*}[tb]
\includegraphics{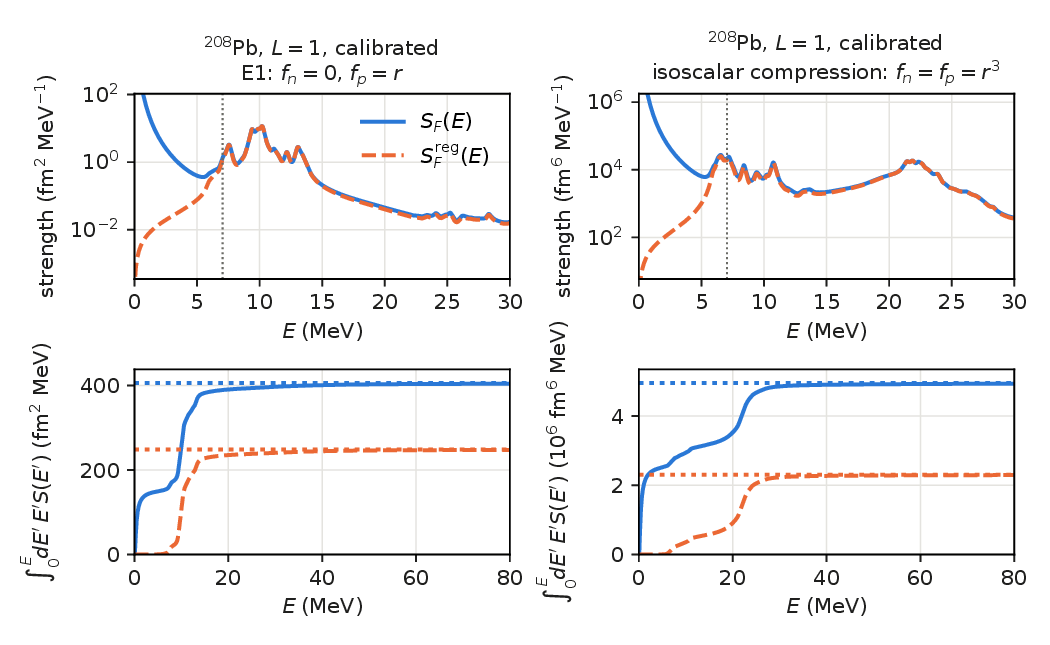}
\caption{$S_F(E)$ (solid) and $S^{\rm reg}_F(E)$ with Eq.~(\ref{eq:Rress}) (dashed) at $\eta=0.2$~MeV for $^{208}$Pb,
$L=1$, in the calibrated Woods--Saxon model ($\Vres$ multiplied by $0.6116$). Left: E1, $f_n=0$, $f_p=r$. Right: isoscalar
compression, $f_n=f_p=r^3$. Top: the vertical dotted line marks the lowest threshold, $7.045$~MeV. Bottom: their energy-weighted sums
$\int_0^EdE'\,E'S(E')$; the dotted lines are their limits,
the right sides of Eqs.~(\ref{eq:m1}) and (\ref{eq:m1res}), $405.6$ and $248.5$~MeV\,fm$^2$ (left), $4.950\times10^6$ and
$2.304\times10^6$~MeV\,fm$^6$ (right). Up to $80$~MeV the sums reach $99.6\%$ of them, above $99.9\%$ with the fraction
$4\eta/(80\pi~\mathrm{MeV})$ that the $E^{-2}$ terms add beyond.}
\label{fig:strength}
\end{figure*}

Answer to Ginocchio et al.:\quad Reference~\cite{GinocchioSchucanWeidenmueller1970} (p.~72) answered that the $S$ matrix
has no pole at $E=0$, because the particle wave functions in the radial integral of its Eq.~(90) have none. They have none, as no $\mathcal H_\alpha$ has a zero-energy bound solution, but in the self-consistent model
that integral, which extends to infinity, diverges as $E^{-2}$. There it reads
$\int_0^\infty dr\,\Psi^0_\alpha(r;E)\,r\,[\tilde\varphi_{q_\alpha}(r)]_\alpha$, with $\Psi^0_\alpha(r;E)$ the scattering solution of $(\mathcal H_\alpha-E)\Psi=0$ in channel $\alpha$ alone. The integrand behaves as
$r\exp\{-[\mathrm{Im}\,k_\alpha(0)-\mathrm{Im}\,k_\alpha(E)]r\}$ for small $E>0$. Two integrations by parts, with
$\mathcal H_\alpha\,r[\tilde\varphi_{q_\alpha}(r)]_\alpha=-(\hbar^2/m)P_{T,\alpha}(r)$ and
$\mathcal H_\alpha P_{T,\alpha}(r)=-\partial_rU^{(q_\alpha)}(r)[\tilde\varphi_{q_\alpha}(r)]_\alpha$ from the hole equation,
give
\begin{equation}
E^2\!\int_0^\infty\!\!dr\,\Psi^0_\alpha r[\tilde\varphi_{q_\alpha}]_\alpha
=\frac{\hbar^2}{m}\!\int_0^\infty\!\!dr\,\Psi^0_\alpha[\tilde\varphi_{q_\alpha}]_\alpha\partial_rU^{(q_\alpha)}.
\label{eq:GinId}
\end{equation}
Its right side is regular at $E=0$, as $\partial_rU^{(q_\alpha)}$ vanishes beyond the radius. At $E=0$, as
$\mathcal H_\alpha\Psi^0_\alpha(r;0)=0$, it is a Wronskian of $\Psi^0_\alpha(r;0)$ with $P_{T,\alpha}(r)$ beyond the
radius, nonzero by Eq.~(\ref{eq:DH}) and (S2).

\emph{Removal by the Mittag-Leffler terms.}\quad As $R_{qq'}(r,r';E)$ is even in $E$, its principal part at the
poles of the mode follows from their states. Since $\tau_x\Lc\tau_x=\Lc$ and $\tau_x\tau_z\tau_x=-\tau_z$,
the matrix $\tau_x\Gc(r,r';E)\tau_x$ solves Eq.~(\ref{eq:Green}) at $-E$, is regular at $r=0$, and decays. Such a
solution is unique at $E$ off the cuts with $\det\J^{(+)}(\pm E)\neq0$, so $\Gc(r,r';-E)=\tau_x\Gc(r,r';E)\tau_x$, and
$\tau_x\xi_q=\xi_q$ makes $R_{qq'}$ even. For a pair $\pm E_g$ of simple poles, each with one state, the residue of $\Gc$ at
$-E_g$ is thus $-\tau_x\psi_g(r)\psi_g^{\Tr}(r')\tau_x$. Subtracting the two pole terms, the principal part of
$R_{qq'}$ at $\pm E_g$, gives
\begin{align}
R^{\rm reg}_{qq'}(r,r';E)={}&R_{qq'}(r,r';E)\nonumber\\
&-\frac{2E_g\,\xi_q^{\Tr}(r)\psi_g(r)\;\xi_{q'}^{\Tr}(r')\psi_g(r')}{E^2-E_g^2}.
\label{eq:Rres}
\end{align}
For the double pole at $E=0$ the even $R_{qq'}$ has no $E^{-1}$ term, and Eq.~(\ref{eq:G2c}) gives the principal part,
its $E^{-2}$ term:
\begin{align}
&R^{\rm reg}_{qq'}(r,r';E)=R_{qq'}(r,r';E)-\frac1{E^2}\lim_{E\to0}E^2R_{qq'}(r,r';E),\nonumber\\
&\lim_{E\to0}E^2R_{qq'}(r,r';E)=\frac{2\,\xi_q^{\Tr}(r)\psi_0(r)\;\xi_{q'}^{\Tr}(r')\psi_0(r')}{\int_0^\infty dr''\,P_0^{\Tr}Q_{P_0}},
\label{eq:Rress}
\end{align}
which is $(4\pi\hbar^2/3mA)\,r^2\partial_r\rho_q(r)\,r'^2\partial_{r'}\rho_{q'}(r')$ in the self-consistent model.

For a real radial field $f_q(r)$, the strength at real $E$ is
$S_F(E)=-\pi^{-1}\mathrm{Im}\sum_{qq'}\int_0^\infty dr\int_0^\infty dr'\,f_q(r)R_{qq'}(r,r';E+i\eta)f_{q'}(r')$,
with the smoothing width $\eta>0$; $S^{\rm reg}_F(E)$ is the same with $R^{\rm reg}_{qq'}$. As $\Lc$ is real,
uniqueness also gives $\Gc(r,r';E^*)=\Gc^*(r,r';E)$ off the cuts, so the subtracted terms are real at real $E$ for the
double pole at $E=0$ and for a pair on the imaginary axis. Hence, as $\eta\to0^+$, $S_F(E)-S^{\rm reg}_F(E)\to0$ at
every real $E\neq0$ in both cases. At $\eta>0$
the double pole leaves in $S_F(E)$ the tail
\begin{align}
&S_F(E)-S^{\rm reg}_F(E)=\frac{2E\eta}{\pi(E^2+\eta^2)^2}\nonumber\\
&\quad\times\sum_{qq'}\int_0^\infty\!dr\int_0^\infty\!dr'\,f_q(r)\lim_{E\to0}E^2R_{qq'}(r,r';E)\,f_{q'}(r'),
\label{eq:tail}
\end{align}
which falls off as $\eta/E^3$ for $E\gg\eta$. In Fig.~\ref{fig:strength} ($\eta=0.2$~MeV) it is $9\%$ of $S_F(E)$ at
$7$~MeV for E1 and $7$--$22\%$ at $6.5$--$10$~MeV for $f_n=f_p=r^3$.

Under (S2), if $\det\J^{(+)}(E)$ has no zero at $\mathrm{Im}\,E>0$, $S_F(E)$ obeys for every $\eta>0$ the energy-weighted sum
rule~\cite{Thouless1961_RPA}, here without self-consistency. With the double pole at $E=0$, Eqs.~(\ref{eq:Rress})
and (\ref{eq:tail}) then give that of $S^{\rm reg}_F(E)$:
\begin{align}
&\int_0^\infty dE\,E\,S_F(E)=m_1=\hbm\sum_q\int_0^\infty dr\,r^2\rho_q(r)\nonumber\\
&\qquad\times\Bigl[\bigl(\partial_rf_q(r)\bigr)^2+\frac{2f_q^2(r)}{r^2}\Bigr],
\label{eq:m1}\\
&\int_0^\infty dE\,E\,S^{\rm reg}_F(E)=m_1-\frac{\bigl[\sum_q\int_0^\infty dr\,f_q\xi_q^{\Tr}\psi_0\bigr]^2}{\int_0^\infty dr\,P_0^{\Tr}Q_{P_0}}.
\label{eq:m1res}
\end{align}
Equation~(\ref{eq:m1}) follows from the $E^{-2}$ term at large $|E|$ of
$\sum_{qq'}\int_0^\infty dr\int_0^\infty dr'\,f_q(r)R_{qq'}(r,r';E)f_{q'}(r')$. Multiplying Eq.~(\ref{eq:Green}) from
the left by $f_q(r)\xi_q^{\Tr}(r)\tau_z/E$, integrating over $r$ by parts without surface terms, and repeating this for
the remainder give no $E^{-1}$ term, since $\xi_q^{\Tr}\tau_z\xi_{q'}=0$, and the coefficient
$\sum_{qq'}\int_0^\infty dr\,f_q\xi_q^{\Tr}\tau_z\Lc\tau_z\xi_{q'}f_{q'}$ of $E^{-2}$. The residual interaction drops
out of it, as velocity-independent contact forces do~\cite{LippariniStringari1989}: the forward and backward parts of
$\tau_z\xi_q$ cancel in the $\Vres$ terms of Eq.~(\ref{eq:Lc}). With the hole equation and (S2), under which the
spin-orbit term of Eq.~(\ref{eq:WS}) cancels, the coefficient is $2m_1$, the ground-state expectation value of the
double commutator of the field with the kinetic energy~\cite{LippariniStringari1989}. The evenness of
$R_{qq'}$ and $R_{qq'}(r,r';E^*)=R^*_{qq'}(r,r';E)$ make $S_F(E)$ odd in $E$, and, with no zero at $\mathrm{Im}\,E>0$,
closing the contour in the upper half plane gives Eq.~(\ref{eq:m1}). In the calibrated
models no zero of $\det\J^{(+)}(E)$ other than $E=0$ is found on the imaginary axis up to $|E|=14$~MeV in steps of
$0.05$~MeV ($^{16}$O) and $0.25$~MeV ($^{208}$Pb). If none lies beyond, none lies at $\mathrm{Im}\,E>0$, as the pair
argument makes $E_g^2$ real.

For the field $f_q(r)=r$, whose $m_1$ a self-consistent model gives entirely to the double pole, half the $E^{-2}$
coefficient of the subtracted terms at large $|E|$ is almost all of $m_1$: $0.950\,m_1$ ($^{16}$O) and $0.903\,m_1$ ($^{208}$Pb) for
the Woods--Saxon pair terms of Eq.~(\ref{eq:Rres}), and $0.915\,m_1$ and $0.924\,m_1$ for the calibrated double pole of
Eq.~(\ref{eq:Rress}). The rest stays in $S^{\rm reg}_F$.

Equality with the projection:\quad For $L=1$ in the self-consistent model with one state at $E=0$, the projection
formula, Eq.~(15) of Ref.~\cite{MizuyamaColo2012_SpuriousSubtraction}, equals the subtraction of Eq.~(\ref{eq:Rress}).
In the radial form of Eq.~(\ref{eq:R}) it is
\begin{align}
\tilde R_{qq'}(r,r';E)={}&R_{qq'}(r,r';E)+\frac{4\pi}{3A}\,r^2\partial_r\rho_q(r)\nonumber\\
&\times\sum_{q''}\int_0^\infty dr''\,r''R_{q''q'}(r'',r';E).
\label{eq:Rtilde}
\end{align}
Multiplying Eq.~(\ref{eq:id1}) from the right by $\xi_{q'}(r')$, with the components
$-(m/\hbar^2)\,r\,[\tilde\varphi_{q_\alpha}(r)]_\alpha$ of $Q_{P_T}$, gives the radial form of the identity of
Ref.~\cite{Vignale1995_sumrule},
\begin{equation}
\sum_{q''}\int_0^\infty dr''\,r''R_{q''q'}(r'',r';E)=-\frac{\hbar^2r'^2\partial_{r'}\rho_{q'}(r')}{mE^2},
\label{eq:dipole}
\end{equation}
so $\tilde R_{qq'}=R^{\rm reg}_{qq'}$. In discrete self-consistent RPA this projection leaves a
remainder that gives no strength for local fields~\cite{Tselyaev2022_SDM}; here it subtracts exactly the principal part,
with continuum boundary conditions.

\emph{Conclusion.}\quad Within the Jost-RPA the spurious c.m.\ mode is a zero of the Jost determinant, an
$S$-matrix pole. In the uncalibrated Woods--Saxon models it is a pair on the imaginary axis. At $E=0$, whether by self-consistency
or by calibration, it is, under (S1) and (S2) and with one state, a pole of exactly second order, with its state from the Autonne--Takagi factorization. In a self-consistent model of one energy density of
$\rho_n$, $\rho_p$ every physical $S$-matrix element continued to $E=0$ carries the $E^{-2}$ term. This corrects only the analytic
conclusion of Ref.~\cite{GinocchioSchucanWeidenmueller1970}; its physical statement that translation is not an intrinsic
excitation stands. Subtracting the principal part of the
response function at the poles leaves the strength at real $E\neq0$ unchanged as $\eta\to0^+$; in such a self-consistent model it equals the projection of
Ref.~\cite{MizuyamaColo2012_SpuriousSubtraction}. If $\det\J^{(+)}(E)$ has no zero at $\mathrm{Im}\,E>0$, the strength
obeys, as in discrete RPA, the energy-weighted sum rule without self-consistency (Fig.~\ref{fig:strength}).

\end{document}